# ACCESS TO DIGITAL PLATFORMS: CAN 'MOBILE' NETWORKS COVERAGE REPORTS BE RELIED UPON? OBSERVATIONS FROM RURAL ZAMBIA AND ZIMBABWE

Gertjan van Stam, Masvingo, Zimbabwe, gertjan@vanstam.net

**Abstract:** As access to digital platforms in Africa is mostly through 'mobile' networks, this paper addresses the mismatch of universalised reports on 'mobile' access and the experiences from digital health practice and cases in Zambia and Zimbabwe. Further, the paper shines a critical light on the meaning of terms like access and the 4$^{th}$ industrial revolution from an African context. It argues for the need to invest in contextual research and development, also to gain a comprehensive understanding of how to access digital platforms in and from Africa.

## 1. INTRODUCTION

In Africa, accessing digital platforms is closely related to the operations of 'mobile' operators as their networks are the only two-way communication network widely available. Digital platforms offer online mechanisms enabling value-creating interactions between external producers and consumers (UNCTAD, 2019, p. 25). 'Mobile' network operators manage fixed infrastructures that consist of a radio access network with base stations, a transmission network providing the communication backbone, and a core network with switching facilities and database servers handling user information. The role out and use of 'mobile' networks is heralded as one of the successes of the introduction of technology in Africa.

The adjective *mobile*, however, is only meaningful in reference to *fixed* phones (sometimes called landlines). In most places in Africa, a fixed, copper-based telephone network has never existed. Therefore, the term *mobile* does not refer to a technological configuration: the mobility or immobility of a particular device (Odumosu, 2018) This ambivalence goes unrecognised in most literature. In this paper, therefore, the term *mobile* is put between quotes or avoided.

The widespread use of communication devices has an enormous influence in Africa. Handheld devices are the contemporary drums of Africa (Nyamnjoh, 2009). Writings on digital developments often start with statistics on the availability of 'mobile' network connectivity and a number of users, gleaned from reports published by the International Telecommunications Union (ITU), the Global System for Mobile Communications Association (GSMA), and other international organisations. In most reports, narratives pivot around the economic benefits of being connected and how connectivity numbers and the potential for expansion represent opportunities for the particular intervention being featured. Here, narratives position telecommunication networks as engines of growth and change, generating benefits and opportunities for the *poor*. However, the categories, definitions, structures, and their assessment appear unaligned with the local experience of connectivity and access, community understandings and the purpose of the appropriation of, and interaction with, digital technologies and services (Loh & Chib, 2021). The reports do not explain the reality of what exists from the perspective of Africans in Africa.

This paper was developed from long-term research into the question "how to regard witnessed mismatches between access realities described in (international) reports and the longitudinal experiences of realities and practices in Zimbabwe and Zambia, among other African countries?" The result is a critique of the implied weight placed on access to digital platforms via 'mobile' telephone networks. It questions the perceptions – *certainties* – derived from reported quantifications.

## 2. METHOD

This research is set along a dynamic and integrative epistemological route, integrating long-term, diverse and differentiated experiences, embodied understandings, value judgements, and actions while residing in rural and urban areas in sub-Saharan Africa (Bigirimana, 2017; van Stam, 2019). The author harmonises longitudinal observations, experiences and learnings of over 19 years living and working in rural areas in Zambia and Zimbabwe, and 2 years in urban areas in Zimbabwe, augmented with most frequent and extensive travels in rural areas in Southern Africa. This is done through reflexive science using the method of living research (van Stam, 2019). The aim is to recognise patterns and wrestle local understanding out from under a Eurocentric gaze as an act of decolonisation (Hlabangane, 2018). As such, the research seeks decentered, inclusive, multifaceted understandings and the emancipation of polyvocality (the consideration of many voices), diversity and multiple perspectives, with preference for the contextual (in this case, African) positionality (Adamu, 2020).

## 3. CONTEXT

The GSMA represents the interests of nearly 800 mobile network operators. The association reports that there were 3.6 billion unique mobile subscribers, half of the world's population, using mobile phones at the end of 2014 (GSMA Intelligence, 2015). In 2015, the GSMA report stated that: "the unconnected population is predominantly rural" (GSMA Intelligence, 2015, p. 3). At the end of 2018, there were 456 million unique mobile users south of the Sahara in Africa (GSMA Intelligence, 2019). This accounts for 44% of the African population south of the Sahara. With regard to the use of digital services, the GSMA reports that 239 million of these use the Internet. The word *rural* is not found in the latter report.

In Zimbabwe, most phone users do not venture beyond the use of basic services, like telephony and instant messaging (Zimbabwe National Statistics Agency (ZIMSTAT) and UNICEF, 2019). In Africa, the use of a phone cannot be readily understood by applying the normative meaning attached to the term *phone* as current in the West. In Africa, 'mobile' phones are to be understood as a concept (Odumosu, 2018). For example, these handheld computing devices act as messaging centres, wallets, and even mirrors. For many, they are an entrenched part of the social constructs of African life. The normative view of the phone's mobility, or as a device for making calls or link to digital platforms, is subject to the fluid and dynamic social function assigned to the technology. For people to connect with digital platforms, such practice must be part of the social constructs of the meaning and function of communication devices in Africa.

The accessibility and use of 'mobile' Internet connectivity vary greatly, depending on the services on offer and the price packaging of so-called *bundles*. In the Africa south of the Sahara, Facebook dominates social media with its WhatsApp service and portal. For instance, early 2020, in Zimbabwe, Facebook traffic was over 85% of all 'mobile' data usage (Postal and Telecommunications Regulatory Authority of Zimbabwe, 2020, p. 10). Having skipped other technologies, and with increasingly powerful computational information and communication devices in the hands of people, 'mobile' phones have led to *leapfrogging* and structure the way many Africans south of the Sahara communicate, handle finances, and are educated (Nyamnjoh, 2009).

In 2015, the Internet Society reported that 94% of the global population was covered by a mobile telecommunication networks, 48% by mobile broadband, and 28% have subscribed to mobile

Internet services. *Coverage*, however, can be defined in many ways. Often, reports are confusing the coverage of population with geographical coverage (land covered). As African cultures closely connect people with notions of geography – *land* – geographical coverage is most important. In Africa, the majority of people live in rural areas. Outside of the main cities, large inequalities can be experienced between the mobile coverage reported and the ability to use mobile services and data-connectivity. Reception of a radio signal (coverage) is often patchy, with large areas without any coverage at all. Witnessing the discrepancy between reports and experienced realities outside of towns makes one wonder about how such reports substantiate their claims of coverage. And, even while a radio signal can be available, data-connectivity can be next to nothing.

Configuring 'mobile' networks in Africa is particular, as users of communication networks behave differently from users elsewhere. Configuring and dimensioning of telecommunication networks in Africa, therefore, diverges most significantly from *the standard rules* as they are set outside of Africa (Odumosu, 2018). For example, as the establishment of communication channels appears to often fail, African users will continue trying to establish the communication channel, even when confronted with a *busy signal*. This African traffic profile has major technical consequences, for instance, for the *signalling traffic* which needs over-dimensioning network backbones and computing systems. With any change of service provisioning, African ways of appropriation and use behaviour have to be understood for dimensioning the technology, so as to ascertain the networks' usefulness to connect with digital platforms.

In the universal reports by the GSMA, Internet Society, and ITU, statements about rural areas are sparse and invariably generic. They often echo an *Africa failing* narrative. Definitions of, or distinctions between, urban and rural areas are commonly absent. Most significantly, the urban/rural disjoins remain unquantified; most access statistics do not differentiate between urban and rural areas and scant (if any) detail is provided on how the numbers break down between them. The lack of detail can also be seen in other fields of interest; for instance, there is no quantitative information on the access of those with disabilities or those among the vulnerable in Africa. Although not made explicit, it seems that available information emerges from desk research, possibly using imputation (Unesco institute for statistics, 2012). Reports seem to reflect the biases induced by the daily experiences of their authors, who appear to be connected and seated in urban areas.

There appears to be great variation – a disconnect – between the macro-figures and practical, local experiences with the availability and usability of access. This disconnect is lost in the arguments derived from those numbers, but obvious to African practitioners, who often struggle to transmit and received digital information (Chawurura, Manhibi, van Dijk, & van Stam, 2020). Questioning these generic reports, some guestimate geographical coverage areas to be as low as 30% in certain countries (Nyambura-Mwaura & Akam, 2013). Even when explicit about geographical coverage, the reliability of the information on *coverage area* from mobile operators is questionable – measurements or imputation? – especially in the large expanses in Africa.

The flattening of the growth of connectivity observed during recent years seem to be turning the tide of generalisation or urban bias. Some initiatives have been taken, such as the ITU resolution in Buenos Aires, shifting the focus towards *inclusion*, with a renewed interest in rural connectivity being featured in a dedicated chapter in its 2020 report (International Telecommunication Union, 2020). Discordant approaches need attention, different from contemporary models that have served for half of the world's population who access digital platforms through the Internet. Obviously, they do not work nor serve the other half.

There are little, if any, dedicated and longitudinal studies on the availability of communication devices and networks utilised to connect with digital platforms from rural areas in Africa. This could be related to the general lack of funding for embedded, African research (Mawere & van Stam, 2019), a general paucity of data (or reluctance to share precious information in situation of power-differences), and Africa's huge size – the second largest continent on Earth. The challenges to physically access and review realities in African rural areas are daunting. Nevertheless, some indications are available, often as a by-product of other research. To substantiate examples of how such information can creatively by recognised, three quantitative cases are presented, one from Zambia and two from Zimbabwe. These cases are gleaned from research on health interventions and digital literacy. Among the research data collected is information on the availability and use of communication devices like phones by clients of rural health institutes.

### 3.1. Early infant diagnosis of HIV in Macha, Zambia (2014–2017)

While reviewing data from a study in early infant diagnosis of HIV exposed children, researchers in Macha, a rural village about 70 km north of Choma in the Southern Province of Zambia, noticed how the availability of phones for *essential use* was quite uncertain (van Dijk et al., 2014). In their study cohort, only 30% of 490 mothers had *used* a phone (reported use in 2013: 33%; 2014: 26%; and 2015: 30%). The researchers argued that, although phone calls and text messaging had the potential to improve early infant diagnosis of HIV exposed children, access issues needed to be addressed first. They recorded low phone ownership, low use, and limited coverage in rural areas around Macha.

### 3.2. Nurse training schools in Masvingo, Zimbabwe (2017)

Jerera is a sizable semi-rural settlement in Masvingo province in southern Zimbabwe, about 100 km south-west of Masvingo town. Musiso Mission hospital is located in Jerera. Silveira is another rural mission settlement with a hospital, also in Masvingo province, 100 km west of Masvingo town. Both hospitals operate a nurse training school (NTS). Musiso's NTS provides a three-year programme for Registered General Nurses. Silveira NTS offers a six-month upgrading programme for Primary Care Nurses.

Upon request by the hospital NTSs, SolidarMed, a Private Voluntary Organisation, established computer labs to support the students and the schools' operations. In the labs, low power, shared computer resources provide access to digitised information. Regular, basic computer training and support is provided by the digital health team of SolidarMed. In 2017, they surveyed the digital literacy among nursing students at these two hospitals (Braat, Sithole, Chikwati, Bishi, & van Dijk, 2018). The survey took place at the start of the school period. The purpose was to establish a baseline on students' knowledge about computers and the use of information and communications technologies (ICT). In total, for the two NTSs, 74 student nurses (of which 62, or 84%, were female) completed a questionnaire. Less than half (31 students, 42%) had ever used a computer before their exposure to the computer lab at the NTS and only 3 (4%) had received training in using computers. About one third of the students (24 students, 32%) owned a computer. Out of 74, only 2 persons (3%) used a computer for their studies: one did an online course and the other visited medical related websites. About a quarter of the students (19 students, 26%) were users of Facebook. The majority of students (56 students, 76%) owned a smartphone. More than half of the respondents (43 students, 58%) did not connect to the Internet on their mobile devices. Less than one third (20 students, 27%) used email (Braat et al., 2018).

### 3.3. The Friendship Bench, Zaka district, Zimbabwe (2018–2019)

In Zimbabwe, the Friendship Bench developed a mental health intervention, aligned with the Zimbabwean context and culture (Chibanda et al., 2016). In 2018, SolidarMed initiated the Friendship Bench method in a closely monitored group of 20 rural health facilities to assess the feasibility of implementation, acceptability and utilisation of the method in rural settings. Operational research accompanied the execution of this work. Binary meta-data regarding the availability of a communication pathway was assessed for 7,230 persons receiving Friendship Bench services in the rural district of Zaka in Masvingo province. Information was gathered from the 20 health facilities during a 13-month period from February 2019 to March 2020. During enrolment, to ensure quality, tracking and tracing, health care clients were asked for a telephone number through which the health institute could reach the client. The availability, or absence, of a telephone number provides an indication of whether or not the person has ready access to means of communications.

Of the 7,230 clients, only 54% provided a telephone number (unpublished data, SolidarMed). In the context where the intervention took place, phone numbers are part of social practices to establish and maintain relationships. Exchanging telephone numbers is regarded as good behaviour. As telephone sharing is common practice (Krah & de Kruijf, 2016), the actual ownership of mobile telephones is likely to be lower than 54% (Okon, 2009). This tallies with the report from the Zimbabwe Multiple Indicator Cluster Survey (MICS), performed between January and April 2019 in 11,091 households across Zimbabwe (Zimbabwe National Statistics Agency (ZIMSTAT) and UNICEF, 2019). The MICS survey found that phone ownership was around 64% in rural areas and around 50% for those categorised poor (Figure 1).

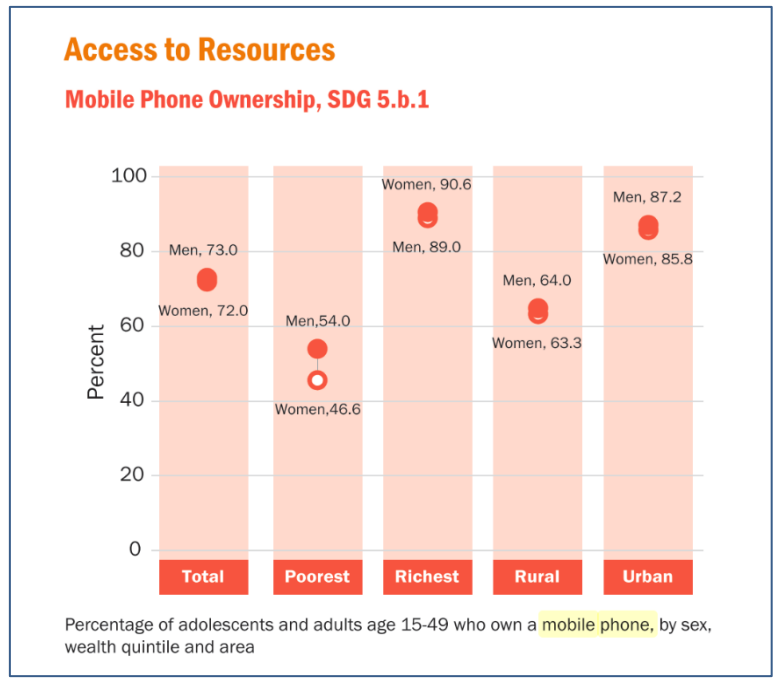

**Figure 1. Phone ownership in Zimbabwe (Zimbabwe National Statistics Agency, 2019)**

## 4. FURTHER OBSERVATIONS

Weak performance of communication networks frustrates the role of digital platforms, services and interventions at health facilities and for clients (Chawurura et al., 2020). In 2018 and 2019, Zimbabwe's Ministry of Health and Child Care endeavoured to decentralise data-entry in the electronic District Health Information System 2 (DHIS2) from district to primary level in Masvingo province, Zimbabwe. The aim was to improve the quality and availability of data and quicken report throughputs. Although around 80% of the 25 facilities involved received a communication network radio signal, implementation of the ideas was severely hampered by poor communication network performance. In two-thirds of the cases, the data throughput was unsuitable for a health information transfer to take place (unpublished data, SolidarMed).

Of course, there are large grey areas regarding the availability, accessibility, and usability of networks. Being in or out of the coverage area of phone networks is an ambiguous judgement. Mobile phone towers can be *out of service* in times of load shedding by the mains electricity supply company, shortages of fuel to run the back-up generator, low solar energy yields, or depleted battery systems. Network design, set up, and management are challenging endeavours in Africa, and implemented technologies might not be conducive for African environments (Johnson & van Stam, 2018) as developments took place without African inputs (van Stam, 2016). And, while data-communications within the communications network can be fine, interconnectivity with other networks can be congested, blocked or managed sub-optimally, while latency can hamper usability of access with digital platforms (Johnson et al., 2012).

In areas of patchy radio coverage, community members often know where a signal can be found. Receiving or making a phone call or sending a message might necessitate climbing a hill or travel to a certain spot. Due to fluctuating propagation of radio waves, network signals are known to fade in and out, especially when the *booster* (telecommunications network tower and equipment) is at a considerable distance or located in a jagged landscape. For instance, at the author's home, 18 km outside of Masvingo town, network access is not possible with a handheld phone: connectivity is achieved by using a high-gain, directional antenna coupled with a router that converts the network signal to an analogue output for a fixed phone and a Wi-Fi signal to access digital platform by means of the Internet Protocol.

## 5. DISCUSSION

In the assessment of access, there is often no clear dichotomy between *haves* and *have-nots*, unless a radio signal or fixed network is not available at all. There is rather a gradation based on different degrees of access to information and communication technologies (Cisler, 2000). A binary divide between the *haves* and *have-nots* is patronising because it fails to value the social resources that diverse groups have (Warschauer, 2003). Therefore, not only is *plain access* crucial, but so is social knowledge and knowledge about how technologies interact in the integration of digital platforms into the practices of communities, institutions, and societies. The aim of information flows should be *social inclusion.* In many parts of Africa, the conventional access model, i.e., one-to-one (or many-to-one) is non-functional, or even potentially destructive (Gurstein, 2012).

The concepts of the *digital divides* and *digital exclusions* are multifaceted and complicated. Definitions remain ambiguous, especially when they are imported from elsewhere. The use of phones in Africa differs from the use of phones in most Western countries, where fixed, mobile, and digital technologies and services are conceptualised as distinct entities, and where they are designed and produced to exist (Alzouma, Chibita, Tettey, & Thompson, 2013). Furthermore, there is no single definition of *rural area*. In the same token, there is no common understanding of the term *poverty*, or consensus on what the boundaries of *availability* are. Concepts of ownership vary widely among cultures, and individual ownership (or individualism per se) is not necessarily a common faculty in Africa (Sheneberger & van Stam, 2011). General assessments of access, as shown above, almost without exception, are set frames derived from the cultures from which the terminologies and subsequent technologies emerge (Mavhunga, 2018; Nkwo, Orji, & Ugah, 2021). Universal definitions are mostly linked to measurable outcomes, devoid of social components like experience, judgement and action. Local perceptions, definitions and interlinked categories denoting *access* may diverge significantly from extant understandings. The use of universal categorisations and their attached boundaries is limited and confusing if their *meaning* is not understandable in context, and have a history of being used to colonise (Mamdani, 2012). Therefore, the definition and interpretation of the availability of access is a complex issue, although crucial when looking at the feasibility of accessing digital platforms.

The fact is that 'mobile' phone networks have been largely designed without contributions from Africa. In the conceptualisation and standardisation of the 5$^{th}$ Generation (5G) mobile networks, no African interest nor embedded knowledge was represented during its constitution (van Stam, 2016). With the installation of imported technical systems, Africa does not primarily purchase technology, but first and foremost establishes links with teams of designers and ways of doing. For instance, after integrating Waze with Google, Noam Bardin (2021) noted "No one buys technology, you buy a team and a way of doing things". These teams and ways of doing increasingly focus on facilitating machine-to-machine communication, instead of communication among people. This shift of attention leads to a reduction in access opportunities, as: "much attention is focused on the roll-out of this the next generation of networks, much of the energy from the mobile sector around connecting the unconnected is dissipating" (Bloom, 2019 online). Alternatives to 'mobile' networks depend on cable, satellite, or stratospheric communication platforms. Sea cables exist in coastal areas, but inland cables are relatively sparse and scarcely reach beyond *the line of rail*. The cost and latencies of satellite systems are often prohibitive for use (Johnson & van Stam, 2018).

In Africa, the 4th industrial revolution – in which automation, like robotics, the industrial Internet and *big* data analytics are positioned as new engines of the economy – is nascent, at best. In most African places, especially rural areas, the second revolution – electrical power – and the third – use of electronics – has not yet taken hold (Naudé, 2017). From the outset, access, among other fundamentals for the 4th industrial revolution, was glaringly lacking (Ministry of Foreign Affairs, 2018; Rodrik, 2018). The pursuit of the 4th industrial revolution could be working against the need to solve the frustrating lack of basic access to digital platforms. A case in point is the lopsided benefits of the digital platform economy. At present, China and the USA own and control around 90% of the worldwide market of digital platforms, while Africa together with Latin America account for only 1% (UNCTAD, 2019). Of course, culturally appropriate alternatives and participatory, community focused and involving arrangements do exist. Examples are the field of community networks (Bidwell, 2020) and dynamic spectrum allocation for the use of TV White Spaces (Gweme & van Stam, 2016).

## 6. CONCLUSIONS

This paper shines a critical light on available, generalised reports on (data) access, as derived from general, mostly international, reference material. In light of the practices and cases presented from Zambia and Zimbabwe, it appears these reports provide *false certainty* and seem to (create/imputate) realities instead of providing a description of realities in rural areas.

Currently, the way in which access and service availability can be understood and used remains obscured by limited research from embedded, African experience and understanding on technology and in rural Africa in general. There is a distinct void in understanding, due to the ongoing normalisation through universalised, external reports. Therefore, the quality and quantity of *access* to digital platforms remains a *shot in the dark* in Africa.

## 7. ACKNOWLEDGEMENT

Thanks to Janneke H. van Dijk and the eHealth team at SolidarMed Zimbabwe for access to their experiences, notes, and helpful suggestions for this paper.

## REFERENCES

Adamu, M. S. (2020). Adopting an African standpoint in HCI4D: A provocation. Conference on Human Factors in Computing Systems - Proceedings. https://doi.org/10.1145/3334480.3382833

Alzouma, G., Chibita, M., Tettey, W., & Thompson, A. (2013). Africa communicating: Digital technologies , representation, and power. Nokoko, 3, 167–177.

Bardin, N. (2021). Why did I leave Google or, why did I stay so long? Retrieved from https://paygo.media/p/25171

Bidwell, N. J. (2020). Wireless in the weather-world and community networks made to last. In PDC'20: Vol.1, June 15-20, Manizales, Colombia (pp. 126–136).

Bigirimana, S. S. J. (2017). Beyond the thinking and doing dichotomy: integrating individual and institutional rationality. Kybernetes, 46(9), 1597–1610.

Bloom, P. (2019). 5G won't reduce the digital divide and might even make it worse. Retrieved June 30, 2020, from https://www.rhizomatica.org/5g-wont-reduce-the-digital-divide-and-might-even-make-it-worse/

Braat, F., Sithole, T., Chikwati, Bishi, J., & van Dijk, J. H. (2018). Evaluation of the utilization of an online perinatal training program. In Conference Presentation at ZiMA Annual Scientific Congress, Harare, Zimbabwe, 15–19 August 2018.

Chawurura, T., Manhibi, R., van Dijk, J. H., & van Stam, G. (2020). Stocktaking the digital health infrastructure in Zimbabwe. In Public Health Conference (ICOPH 2020).

Chibanda, D., Weiss, H. A., Verhey, R., Simms, V., Munjoma, R., Rusakaniko, S., … Araya, R. (2016). Effect of a primary care–based psychological intervention on symptoms of common mental disorders in Zimbabwe. A randomized clinical trial. Jama, 316(24), 2618–2626. https://doi.org/10.1001/jama.2016.19102

Cisler, S. (2000). Subtract the "Digital Divide." Retrieved July 1, 2020, from http://www.athenaalliance.org/rpapers/cisler.html

GSMA Intelligence. (2015). The Mobile Economy. London: GSMA.

GSMA Intelligence. (2019). The Mobile Economy. Sub-Saharan Africa 2019. London: GSMA.

Gurstein, M. (2012). The mobile revolution and the rise of possessive individualism. Retrieved July 1, 2020, from http://gurstein.wordpress.com/2012/07/21/the-mobile-revolution-and-the-rise-and-rise-of-possessive-individualism/

Gweme, F., & van Stam, G. (2016). The potential for use of TV White Spaces for the internet in Zimbabwe. In 1st Institute of Lifelong Learning and Development Studies International Research Conference, Chinhoyi University of Technology, 2-5 August 2016, Chinhoyi, Zimbabwe. Retrieved from https://www.researchgate.net/publication/305807591_The_Potential_for_use_of_TV_White_Spaces_for_the_Internet_in_Zimbabwe

International Telecommunication Union. (2020). Measuring digital development. Facts and figures, 2020. Retrieved from https://www.itu.int/en/mediacentre/Documents/MediaRelations/ITU Facts and Figures 2019 - Embargoed 5 November 1200 CET.pdf

Johnson, D. L., Mudenda, C., Pejovic, V., Sinzala, A., van Greunen, D., & van Stam, G. (2012). Constraints for Information and Communications Technologies implementation in rural Zambia. In K. Jonas, I. A. Rai, & M. Tchuente (Eds.), e-Infrastructure and e-Services for Developing Countries. AFRICOMM 2012. Lecture Notes of the Institute for Computer Sciences, Social Informatics and Telecommunications Engineering, vol 119. Berlin, Heidelberg: Springer.

Johnson, D. L., & van Stam, G. (2018). The shortcomings of globalised internet technology in southern Africa. In T. F. Bissyandé & O. Sie (Eds.), e-Infrastructure and e-Services for Developing Countries. AFRICOMM 2016. Lecture Notes of the Institute for Computer Sciences, Social Informatics and Telecommunications Engineering, vol 208. Cham: Springer.

Krah, E. F., & de Kruijf, J. G. (2016). Exploring the ambivalent evidence base of mobile health (mHealth): A systematic literature review on the use of mobile phones for the improvement of community health in Africa. Digital Health, 2. https://doi.org/10.1177/2055207616679264

Loh, Y. A.-C., & Chib, A. (2021). Reconsidering the digital divide: An analytical framework from access to appropriation. Information Technology and People. https://doi.org/10.1108/ITP-09-2019-0505

Mamdani, M. (2012). Define and rule: Native as political identity (Kindle). Cambridge: Harvard University Press.

Mavhunga, C. C. (Ed.). (2018). What do science, technology, and innovation mean from Africa? Cambridge: MIT Press. https://doi.org/10.7551/mitpress/9780262535021.001.0001

Mawere, M., & van Stam, G. (2019). Research in Africa for Africa? Probing the effect and credibility of research done by foreigners for Africa. In P. Nielsen & H. C. Kimaro (Eds.), Information and Communication Technologies for Development. Strengthening Southern-Driven Cooperation as a Catalyst for ICT4D. ICT4D 2019. IFIP Advances in Information and Communication Technology, vol 552 (pp. 168–179). Cham: Springer.

Ministry of Foreign Affairs. (2018). Transition and inclusive development in Sub-Saharan Africa: An analysis of poverty and inequality in the context of transition. The Hague: Ministry of Foreign Affairs.

Naudé, W. (2017). The fourth industrial revolution in Africa: potential for inclusive growth? Retrieved June 30, 2020, from https://www.thebrokeronline.eu/the-fourth-industrial-revolution-in-africa-potential-for-inclusive-growth/

Nkwo, M. S., Orji, R., & Ugah, J. (2021). Insider perspectives of human-computer interaction for development research: opportunities and challenges. In AfriCHI 2021, 10-12 March 2021, online.

Nyambura-Mwaura, H., & Akam, S. (2013). Telecoms boom leaves rural Africa behind. Retrieved December 14, 2020, from http://www.reuters.com/article/2013/01/31/us-africa-telecoms-idUSBRE90U0MK20130131

Nyamnjoh, F. B. (2009). Mobile phones: The new talking drums of everyday Africa. (M. de Bruijn, F. B. Nyamnjoh, & I. Brinkman, Eds.). Bamenda: Langaa RPCIG.

Odumosu, T. (2018). Making mobiles African. In C. C. Mavhunga (Ed.), What do science, technology, and innovation mean from Africa? (pp. 137–150). Cambridge: MIT Press.

Okon, U. A. (2009). Information communication technology and sustainable communities in Africa: The case of the Niger Delta region of Nigeria (Feb 2009). In Information and Communication Technologies and Development (ICTD 2009) (pp. 367–378). https://doi.org/10.1109/ICTD.2009.5426690

Postal and Telecommunications Regulatory Authority of Zimbabwe. (2020). Abridged postal & telecommunications sector performance report, first quarter 2020. Harare: PORTRAZ.

Rodrik, D. (2018). An African growth miracle? Journal of African Economies, 27(1), 10–27. https://doi.org/10.1093/jae/ejw027

Sheneberger, K., & van Stam, G. (2011). Relatio: An examination of the relational dimension of resource allocation. Economics and Finance Review, 1(4), 26–33.

UNCTAD. (2019). Digital economy report 2019: Value creation and capture - implications for developing countries. New York: United Nations.

Unesco institute for statistics. (2012). UIS frequently asked questions. Retrieved from http://uis.unesco.org/sites/default/files/documents/uis-frequently-asked-questions-education-statistics-2016-en.pdf

van Dijk, J. H., Moss, W. J., Munsanje, B., Sinywimaanzi, C., Thuma, P. E., & Sutcliffe, C. G. (2014). Feasibility of using mHealth to improve early infant diagnosis of human immunodeficiency virus infection in rural southern Zambia. In 11th Prato CIRN Conference, 13-15 Oct 2014, Prato, Italy.

van Stam, G. (2016). Africa's non-inclusion in defining fifth generation mobile networks. In T. F. Bissyande & O. Sie (Eds.), e-Infrastructure and e-Services for Developing Countries. AFRICOMM 2016. Lecture Notes of the Institute for Computer Sciences, Social Informatics and Telecommunications Engineering, vol 208. Cham: Springer.

van Stam, G. (2019). Why do foreign solutions not work in Africa? Recognising alternate epistemologies. In M. van Reisen, M. Mawere, M. Stokmans, & K. A. Gebre-Egziabher (Eds.), Roaming Africa: Migration, Resilience and Social Protection (pp. 55–82). Bamenda: Langaa RPCIG.

Warschauer, M. (2003). Technology and social inclusion. Rethinking the digital divide. Cambridge: MIT Press.

Zimbabwe National Statistics Agency. (2019). Zimbabwe Multiple Indicator Cluster Survey 2019, snapshots of key findings. Harare: ZIMSTAT.

Zimbabwe National Statistics Agency (ZIMSTAT) and UNICEF. (2019). Multiple Indicator Cluster Survey (MICS) 2019, survey findings report. Harare: ZIMSTAT and UNICEF.